%Paper: gr-qc/9304017
%From: Piotr Koc <ufkoc@ztc386a.if.uj.edu.pl>
%Date: Wed, 14 Apr 93 11:48:18 cet

%THIS IS  emTeX Version 3.0 [3a].
\magnification=\magstep1
\centerline{\sl CONDENSATION OF MATTER AND TRAPPED SURFACES}
\centerline{\sl IN QUASI-POLAR GAUGE}
\vskip 1cm
\centerline{\bf Piotr Koc\footnote{*}
{\rm e-mail address: \tt ufkoc@ztc386a.if.uj.edu.pl}}
\vskip 1cm
\centerline{\bf Institute of Physics}
\centerline{\bf Jagellonian University}
\centerline{\bf Reymonta 4, 30-059 Krak\'ow,}
\centerline{\bf Poland}
\vskip 1cm
\centerline{PACS numbers: 04.20.Cv, 97.60.Lf}
\vskip 1cm
\centerline{\bf Abstract}
We continue the investigation of formation of trapped surfaces in strongly
curved , conformally flat geometries. Initial data in quasi-polar gauges
rather then maximal ones are considered. This implies that apparent horizons
coincide with minimal surfaces. Necessary and sufficient conditions for the
formation of trapped surfaces are given. These results can be generalized
to include the case with gravitational radiation. We found that mass of a
body inside a fixed volume is bounded from above if geometry of a Cauchy
slice is smooth.
\vskip 1cm
\centerline{\hrulefill}
\vskip 1cm
\par
{\bf 1.INTRODUCTION}
\par
There is a widespread opinion that matter squeezed in a small
enough volume must collapse under its own weight. This idea may be
expressed as the so called {\bf hoop conjecture} which
states [1]: "An event horizon will form if and only if $4\pi m>C$, where $C$
is the smallest hoop that can be passed around body in any direction". In
the original formulation, $m$ is the ADM mass. It is known that
this formulation understood as strict criterion fails [2]; that means that mass
must be concentrated in all three direction. The most significant progress was
made in proving {\bf the trapped surface conjecture} which says that:
"Any mass that is concentrated in a region of sufficiently small
diameter can be surrounded by a trapped surface" [3]. If we assume Cosmic
Censorship, {\bf the existence of a trapped surface (even on an initial
slice) implies the existence of black
holes}, as suggested by the Penrose-Hawking singularity theorems [4].
In this paper we will examine formation of trapped surfaces in
a class of nonspherical initial data sets.
\par
This problem may be also seen from another point of view - as a part of the
proof of Cosmic Censorship Hypothesis [5].
\par
 The trapped surface conjecture and the problem of maximal concentration of
matter have been solved recently in series of papers (see [6] for review)
for different classes of geometries. All this work was made in
{\bf the maximal slicing} gauge (or even under a more restrictive assumption of
time-symmetry initial data):
$$tr K=0.$$
$K$ is an extrinsic curvature tensor of a Cauchy slice. There
exists another attractive gauge (the polar gauge)
for spherically symmetric Cauchy data, in which
$tr K=K^r{}_r$, where $r$ is the radial coordinate on the Cauchy surface. As
will be pointed out later, in the polar gauge the apparent horizons coincide
with minimal surfaces; that enormously simplifies the formulation of criteria
for the formation of trapped surfaces. In this paper we investigate the
formation of trapped surfaces in {\bf the generalized polar gauge} in which
the spacetime is foliated by hypersurfaces such that
$$tr K=K^\sigma{}_\sigma.\eqno(1.1)$$
Subscript $\sigma$ is connected with a quasi-radial coordinate (see below for
more detailed description); there is no summation over $\sigma$.
Such foliations were considered in [7].

\par
This paper is organized as follows. Section 2 is introductory one.
In Section 3 it is proven that if total rest mass exceeds certain
critical value determined by a size of the system then averaged trapped surface
must exist (sufficient condition). Section 4 shows, that there is an
upper limit  for the amount of matter inside a volume of a fixed size (an
analogous result was proven earlier [6,8] in the maximal slicing).
A necessary condition for the formation of averaged trapped surfaces
(and thus, also of pointwise trapped surfaces)
is analyzed in the case of spherical symmetry in Section 5.
We would like to point out that the quasi-polar gauge allows one to obtain a
necessary condition even for non-symmetric in time initial data, in contrast
with maximal gauges. Necessary and sufficient
condition stated in Secs. 3, 5 have been proven earlier [6]
but under assumption of momentarily static initial data.
Section 6 presents a general criterion for the existence of a trapped surface
in a system possibly containing the gravitational radiation. The considered
self gravitating system is not required to be conformally flat. The last
Section comprises some comments on the obtained results.

\vskip 1cm
\par
{\bf 2.PRELIMINARIES}
\par
Let us start from the initial value equations:
$$^{(3)}R[g]-K_{ab}K^{ab}+(trK)^2=16\pi\rho,\eqno(2.1) $$
$$D_aK^a{}_b-D_bK^a{}_a=-8\pi J_b.\eqno(2.2)$$
The notation of this paper is taken from the first reference of [9].
Energy density $\rho$ and matter current $J_b$ are prescribed on a space-like
hypersurface $\Sigma$. Three-geometry of $\Sigma$ is defined by the induced
metric tensor $g_{ab}$. An immersion into a 4-dimensional space-time is
given by an extrinsic curvature tensor $K_{ab}={1\over2}{\cal L}_tg_{ab}.$
In (2.1) $^{(3)}R[g]$ is the scalar curvature of $\Sigma$.
$D_a$ in (2.2) denotes covariant derivative in the 3-dimensional metric.
We put c=1, G=1.
\par
We assume a conformally flat geometry,
$$g_{ab}=\Phi^4\widetilde g_{ab},$$
in the next 4 sections. $\Phi$ is a conformal
factor and $\widetilde g_{ab}$ denotes a flat metric. All quantities with tilde
refer to flat background space.
Assumption of the conformal flatness may be intuitively understood as
a restriction to the case without gravitational waves - a conformally flat
system without matter is flat.
It allows us to use the total proper rest mass as the quasilocal
measure of the gravitational energy.
\par
Now we employ, special coordinates [10] adapted to equipotential surfaces $S$
(i.e., surfaces of constant $\Phi$). In these coordinates the 3-dimensional
line element has a form
$$ds^2=\Phi^4(\sigma)(\widetilde g_{\sigma\sigma}d\sigma^2+
\widetilde g_{
\tau\tau}d\tau^2+\widetilde g_{\phi\phi}d\phi^2
+ 2\widetilde g_{\tau \phi }d\tau d\phi).$$
Here $\sigma\geq 0,$ $\sigma$ foliates the level surfaces of
the $\Phi$, that from now one are {\bf assumed to be convex},
and $\tau,\phi$ are quasi-angle variables.
The scalar curvature of the conformally flat metric $g_{ab}$ reads
$$^{(3)}R[g]=-8{\nabla ^2\Phi\over{\Phi ^5}}.$$
Inserting this into (2.1) we obtain the following equation:
$$\nabla ^2\Phi+{\Phi^5\over 8}[K_{ab}K^{ab}-(trK)^2] =
-2\pi\rho\Phi ^5.\eqno(2.3)$$
\par
A surface $S$ is called {\bf an averaged trapped surface} [10], if
the mean expansion $\vartheta$ of outgoing future directed
null geodesics which are orthogonal to $S$ is nonpositive,
$$\int\limits_S\vartheta dS\leq 0.$$
$\vartheta$ is related to the initial data of Einstein
equations $g_{ab}, K_{ab}$ by
$$\vartheta=D_an^a-K_{ab}n^a n^b+trK,$$
where $n^a$ denotes a normal unit vector to $S$.
The foliation condition (1.1) implies, that the last two terms in the above
equation cancel and $\vartheta=D_an^a$. Thus we can rewrite the condition
for an averaged trapped surface in the form
$$\int\limits_{\widetilde S}D_an^a d\widetilde S\leq 0.\eqno(2.4)$$
A closed surface $S\subset\Sigma$ is called minimal if $D_an^a=0$ everywhere
on $S$ and it is called an apparent horizon if $\vartheta (S)=0$. Thus on
a quasi-polar Cauchy slice $\Sigma$ (on which $tr K=K^{\sigma}{}_{\sigma}$)
the notions of apparent horizons and minimal surfaces coincide. This means
that the existence problem of apparent horizons becomes an intrinsic problem
of a 3-dim. riemanian geometry.

\vskip 1cm
\par
{\bf 3.SUFFICIENT CONDITION}
\par
In this chapter we find a sufficient condition for an equipotential surface
$S$ to be an averaged trapped surface. Mass contained in a volume $V$ bounded
by a surface $S$ is defined by
$$M=\int\limits_V\rho dV=
\int\limits_{\widetilde V}\rho\Phi ^6
 d\widetilde V.\eqno(3.1)$$
Let us multiply equation (2.3) by $\Phi$ and integrate over the volume
$\widetilde V$. Using the above definition of $M$ we obtain
$$M=-{1\over 2\pi}\int\limits_{\widetilde V}
\Phi\nabla ^2\Phi d\widetilde V
-{1\over 16\pi}\int\limits_{\widetilde V}[K_{ab}K^{ab}-(trK)^2]\Phi^6
d\widetilde V=$$
$$={1\over 2\pi}\int\limits_{\widetilde V}
\partial_i\Phi\partial^i\Phi d\widetilde V
-{1\over 2\pi}\int
\limits_{\widetilde S}\Phi\partial_{\sigma}\Phi {1\over
\sqrt{\widetilde g_{\sigma\sigma}}}d\widetilde S
-{1\over 16\pi}\int\limits_{\widetilde V}[K_{ab}K^{ab}-(trK)^2]\Phi^6
d\widetilde V.\eqno(3.2)$$
The divergence of the normal unit vector $n^i$, i.e.,
the mean curvature of a surface $S$ in the physical metric $g_{ik}$
which defines trapped surface (eq.(2.4)) has the following form
$$D_i n^i={1\over\sqrt g}
\partial_i(\sqrt g n^i)={{\partial_{\sigma}ln\sqrt
{\xi}}\over\Phi^2
\sqrt{\widetilde g_{\sigma\sigma}}}+{4\Phi \partial_{\sigma}
\Phi \over{\Phi^4\sqrt
{\widetilde g_{\sigma\sigma}}}},\eqno(3.3)$$
where
$$\xi=\widetilde g_{\tau\tau}\widetilde g_{\phi\phi}-
(\widetilde g_{\tau\phi})^2.$$
We also need to calculate the mean curvature of $S$ with respect
to the background flat metric $\widetilde g_{ik}$
$$\widetilde p={1\over\sqrt {\widetilde g}}\partial_i(\sqrt{\widetilde g}
\widetilde n^i)={\partial_{\sigma}ln\sqrt{\xi}\over\sqrt{\widetilde g
_{\sigma\sigma}}},\eqno(3.4)$$
By inserting (3.4) to (3.3) we get
$$\Phi\partial_{\sigma}\Phi=
\Bigl (D_i n^i-{\widetilde p\over\Phi^2}\Bigr ){\Phi^4
\sqrt{\widetilde g_{\sigma\sigma}}\over4}.\eqno(3.5)$$
In (3.5) we recognize the integrand of the second term of (3.2) hence (3.2)
reads now
$$M={1\over 2\pi}
\int\limits_{\widetilde V}\partial_i\Phi\partial^i\Phi
d\widetilde V-{\Phi^4 \over 8\pi}
\int\limits_{\widetilde S}D_i n^i d\widetilde S+
{\Phi^2\over 8\pi}
\int\limits_{\widetilde S}\widetilde p d
\widetilde S+$$
$$-{1\over 16\pi}\int\limits_{\widetilde V}[K_{ab}K^{ab}-(trK)^2]\Phi^6
d\widetilde V\eqno(3.6)$$
Now let us assume that S {\bf is not an averaged trapped} surface, that is
$$\int\limits_{\widetilde S}D_i n^i d\widetilde S>0.$$
In such a case the second term of (3.6) is negative. The gauge condition
(1.1) implies
$$K_{ab}K^{ab}-(trK)^2=K_{ab}K^{ab}-K_{\sigma\sigma}K^{\sigma\sigma}\geq 0;
\eqno(3.7)$$
therefore the last term of (3.6) is also nonpositive. Therefore, if $S$ is not
an averaged trapped surface then the following inequality holds:
$$M<{1\over 2\pi}\int\limits_{\widetilde V}
\partial_i\Phi\partial^i\Phi d\widetilde V+{\Phi^2\over 8\pi}\int\limits
_{\widetilde S}\widetilde p d\widetilde S.\eqno(3.8)$$
Let us make two definitions
$$D(S(\sigma))={1\over 2\pi}\int\limits_{\widetilde V}\partial_i\Phi\partial^i
\Phi d\widetilde V+{\Phi^2\over\ 8\pi}\int\limits_{\widetilde S}
\widetilde p d\widetilde S,\eqno(3.9)$$
$$Rad(S(\sigma ))=D(S(0))+{1\over 8\pi}\int\limits_0^\sigma\Phi^2(s)
\partial_s\biggl (\int\limits_{\widetilde S} \widetilde p d\widetilde S\biggr )
ds.\eqno(3.10)$$
\par
{\bf Remark.} Let us comment on the interpretation of $Rad(S(\sigma))$.
In the case of spherical symmetry
$D(S(0))=\lim\limits_{\sigma\rightarrow 0} D(S(\sigma))=0$.
Mean curvature $\widetilde p$ of a sphere
equals to $2/r$, hence $Rad(S(\sigma ))$ reduces to the proper radius
$L=\int\limits_0^R \Phi^2 dr$.
\par
It is reasonable to conjecture, that $Rad(S(\sigma))$ is
bounded from above by the largest proper radius of the
volume enclosed by the surface $S$ in all
geometries when the equipotential surfaces are convex
[10,11] (that holds true in spheroidal geometries).
\par
Using the definition of
$D(S(\sigma))$ we can rewrite (3.8) in the form
$$M<D(S(\sigma)).$$
Taking into account (2.3, 3.7) and {\bf assuming positivity of the
energy density} $\rho\geq 0$ we get $\nabla^2\Phi\leq 0$. From the
maximum principle we have
$$\partial_\sigma \Phi\leq 0.\eqno(3.11)$$
This allows us to use the following estimation
$$D(S(\sigma))\leq Rad(S(\sigma)),\eqno(3.12)$$
which was proved in [9]. Finally, we obtain
$$M<Rad(S(\sigma));$$
in summary, if $S$ is not trapped, then the amount of mass inside it can not
exceed $Rad(S(\sigma))$. Thus by contradiction we have proven the following:
\par
{\bf Theorem 1.} (sufficient condition). Assume a
quasi-polar Cauchy slice $tr K=K^\sigma{}_\sigma$, with conformally flat
metric and nonnegative energy density $\rho\geq 0$.
If the amount of mass inside $S$ satisfies the inequality
$$M\geq Rad(S(\sigma)),$$
then a convex equipotential surface $S$ is trapped.
\par
Let me remark, that actually the form of the sufficient condition does not
depend on the fact if matter is moving or not. Let us point out, that is
unlike situation studied hitherto in spherically symmetric geometry [12],
where coefficients were different in these two situations.

\vskip 1cm
\par
{\bf 4.MASS OF A BODY IS BOUNDED}
\par
Using eq. (3.5) we get
$$D_i n^i={\widetilde p\over\Phi^2}+{4\partial_\sigma\Phi
\over\Phi^3\sqrt{\widetilde g_{\sigma\sigma}}}.$$
Inserting this equation in (3.6) and replacing
the first and the third terms of (3.6)
by $D(S(\sigma))$ (see eq. (3.9)) we obtain
$$M=D(S(\sigma))-{\Phi\over 4\pi}\int\limits_{\widetilde S}
\Bigl ({1\over 2}\Phi\widetilde p+2\widetilde n^\sigma\partial_
\sigma\Phi\Bigr )d\widetilde S
-{1\over 16\pi}\int\limits_{\widetilde V}[K_{ab}K^{ab}-(trK)^2]\Phi^6
d\widetilde V.$$
The above formula can be rewritten in following form
$$M=D(S(\sigma))+{\Phi^2\over 8\pi}\int\limits_{\widetilde S}
\widetilde p d\widetilde S-{\Phi\over 4\pi}\int\limits_
{\widetilde S}(\Phi\widetilde p+2\widetilde n^\sigma\partial_\sigma
\Phi)d\widetilde S+$$
$$-{1\over 16\pi}\int\limits_{\widetilde V}[K_{ab}K^{ab}-(trK)^2]\Phi^6
d\widetilde V$$
In the case of decreasing $\Phi$ (eq.(3.11)) and assuming convexity of $S$,
the following technical estimation has been found [9]:
$$\int\limits_{\widetilde S}
(\Phi\widetilde p+2\widetilde n^\sigma\partial_\sigma\Phi)
d\widetilde S\geq 0.$$
{}From the above inequality and (3.7) we see that last two terms are
nonpositive.
Using definition (3.9) of $D(S(\sigma))$ we conclude that
$$M\leq 2D(S(\sigma)).$$
Because of (3.12) the following holds true.
\par
{\bf Theorem 2.} Under the assumptions:
$$\rho\geq 0,$$
$$tr K=K^\sigma{}_\sigma$$
the amount of mass inside a convex equipotential surface $S$ is bounded from
above,
$$M\leq 2Rad(S(\sigma)).$$

\vskip 1cm
\par
{\bf 5.NECESSARY CONDITION}
\par
Now we restrict ourselves to the {\bf case of spherical symmetry}.
Then the extrinsic curvature tensor can be written:
$$K_{ab}=K(r)n_a n_b+H(r)g_{ab}.$$
Taking into account our foliation condition $trK=K^r{}_r$  it is easy to
see that $K_{ab}$ simplifies to
$$K_{ab}=K(r)n_a n_b.$$
In such a case the term containing extrinsic curvature vanishes,
$K_{ab}K^{ab}-(trK)^2=0$. Now eq.(3.2) reads
$$M=-2R^2\Phi(R)\Phi'(R)+2\int\limits_0^R\Phi'^2r^2dr.\eqno(5.1)$$
Let us assume, that surface $S$ {\bf is trapped}: $\vartheta=D_a n^a\leq 0$.
It implies, that $2\Phi'R+\Phi\leq 0$. Using this inequality and (5.1) we
obtain
$$M\geq R\Phi(R)^2+2\int\limits_0^R\Phi'^2r^2dr.\eqno(5.2)$$
The right hand side of (5.2) is bounded from below by $L/2$ [12], therefore we
may
conclude the following
\par
{\bf Theorem 3.} Assume that a sphere $S$ is trapped and $tr K=K^r{}_r$, then
$$M\geq{L\over 2}.$$
\par
Let us remark, that the last estimation holds true also for moving matter,
while an analogous result of [12] has been proven
only for momentarily static initial data. The use of quasi-polar Cauchy
slices allows one to get more information than the use of maximal slices.
\par
Let us consider explicit example. Take a spherically symmetric
shell of a coordinate radius $R$: $\rho=\rho_0\delta (r-R)$. Due to the
vanishing of term with an extrinsic curvature the solution of the constraint
(2.3) has a form identical with a solution in the maximal gauge (see[8]):
$$\Phi (r)=1+{m\over 2R}\qquad for\qquad r\leq R,$$
$$\Phi (r)=1+{m\over 2r}\qquad for\qquad r > R,$$
where $m$ denotes ADM mass of a system. Trapped surfaces exist if $m\geq 2R$.
The surface matter density $\rho_0$ is related to $m$ by
$$\rho_0={m\over4\pi R^2(1+{m\over 2R})^5}.$$
By changing $\rho_0$ we may increase $m$ to a large enough value which
satisfies the inequality $m\geq 2R$ and then allows one to form a trapped
surface. The polar gauge {\bf does not} prevent the formation of
trapped surfaces {\bf on the initial Cauchy surface}.

\vskip 1cm
\par
{\bf 6.GENERALIZATION}
\par
Now we resign from the assumption of the conformal flatness. Let $U[S(\sigma)]$
be a foliation of Cauchy hypersurfaces by closed 2-surfaces; $\sigma$ is a
parameter, that enumerates leaves of the foliation.
In [13] following lemma has been proven:
$$\partial_{\sigma}\int\limits_{S(\sigma)}D_in^idS=\int\limits_{S(\sigma)}
\bigl (2 det(D_i n^k)-R^{\sigma}{}_{\sigma}\bigr )n_{\sigma} dS.\eqno(6.1)$$
The scalar curvature can be expressed in terms of sectional and Ricci
curvatures by identity
$$^{(3)}R[g]=2(R^{\tau\phi}_{\tau\phi}+R^{\sigma}{}_{\sigma}),$$
where $\tau,\phi$ denotes two different direction tangent to a leaf of
$S(\sigma)$.
Inserting $R^{\sigma}{}_{\sigma}$ from the above equation and using the well
known Gauss formula ($K$ denotes Gauss curvature)
$$det(D_i n^k)=K-R^{\tau\phi}_{\tau\phi}$$
into (6.1) we obtain
$$\partial_{\sigma}\int\limits_{S(\sigma)}D_in^idS=\int\limits_{S(\sigma)}
\Bigl (2K-{^{(3)}R\over2}-R^{\tau\phi}_{\tau\phi}\Bigr )n_{\sigma}dS.$$
Integrating this with respect to $\sigma$ and using the constraint equation
(2.1) we get
$$\int\limits_{S(\rho)}D_in^i dS=\int\limits_{S(0)}
D_i n^i dS+\int\limits_0^{\rho} d\sigma\int\limits_{S(\sigma)}2Kn_{\sigma}dS
-8\pi M(S(\rho))+$$
$$-\int\limits_{V(\rho)}\Bigl
(R^{\tau\phi}_{\tau\phi}+{K_{ab}K^{ab}-(trK)^2\over2}\Bigr
)dV.\eqno(6.2)$$
Size of a configuration may be defined by [13]
$$RAD(S(\rho))={1\over 8\pi}\int\limits_0^{\rho}d\sigma\int\limits_{S(\sigma)}
2Kn_{\sigma}dS.$$
This quantity coincides with $Rad(S(\rho))$ in a conformally flat case and
in the spherically symmetric case it is equal to a proper radius of
a configuration. Let us define
$$E(S(\rho))={1\over 8\pi}\int\limits_{V(\rho)}\Bigl
(R^{\tau\phi}_{\tau\phi}+{(K^{\tau}
{}_{\tau})^2+(K^{\phi}{}_{\phi})^2+2(K^{\tau}{}_{\phi})^2\over 2}\Bigr )dV,$$
and call $E(S)$ the radiation energy . For more details on $E(S)$ see [13]
where it was defined in a slightly different way. The left hand side of
(6.2) is the mean expansion which defines a averaged trapped surface
(eq. (2.4)).
Let we assume that the foliation is smooth and locally convex. It implies that
$\int_{S(0)}D_in^idS=0$.
Taking into account the above definitions of $RAD(S(\rho))$ and $E(S(\rho))$
we infer from (6.2) the following:
\par
{\bf Theorem 4.} The surface $S$ is average trapped if and only if
$$M(S)+E(S)\geq RAD(S).$$

\vskip 1cm
\par
{\bf 7.SUMMARY}
\par
We have considered a quasi-polar Cauchy slice, i.e. a Cauchy hypersurface
satisfying the quasi-polar gauge condition with a conformally flat and convex
(in the sense defined above) three-geometry. In such a case the system
of a fixed size $Rad(S)$ must have rest mass $M(S)$ smaller then
$2Rad(S)$ (theorem 2). Theorem 1 gives a sufficient condition for an
equipotential surface $S$ to be an average trapped surface. The "only if" part
of the trapped surface conjecture  is formulated, in the case of spherical
symmetry,
in theorem 3. Only theorems 1 and 2 require the nonnegativity of the
energy density. We would like to stress out that all result obtained here are
strict. It is easy to see that all inequalities will be saturated in
the spherically symmetric or/and  momentarily static initial slices.
We should point out that methods applied in this paper were used earlier in
the case of maximal slices. The use of quasi-polar slices, however, simplifies
the whole analysis and makes it possible to draw more conclusive
statements than  in the case of maximal Cauchy slices. In particular, it
allows one to cover the case with nonzero matter fluxes on the Cauchy surface.
\par
As it was pointed out before, the existence problem of polar gauge foliation is
not
solved yet, similarly as for maximal and constant extrinsic curvature
foliations. There is known a result which suggests
that the latter foliations do not cover all physical spacetimes. D.M. Eardley
and L.Smarr [14] showed that foliations $trK=0$ and $trK=const$ avoid
some kinds of singularities. We suppose that this also may take a place in
the polar (quasi-polar) gauge. The problem of the efficiency of gauges is of
utmost importance but it is out of the scope of the present paper.
\par
{\bf Acknowledgements.} The author thanks Dr. Edward Malec for motivating
discussions. This work is supported by the Polish Government Grant no
{\bf PB 2526/92} and {\it Stefan Wajs} foundation.

\vskip 1cm
\par
\centerline{\bf REFERENCES}
\par
[1] K.Thorn in:"Magic without magic: John Archibald Wheeler"; J.Klauder (ed.)
(Freeman, San Francisco, 1972) p.231; C.W.Misner, K.Thorn, J.Whiler,
Gravitation (Freeman, San Francisco 1973) p.867.\par
[2] W.B.Bonnor, Phys. Lett. 99A, 424(1983).\par
[3] H.J.Seifert, Gen.Rel.Grav.10,1065(1976).\par
[4] S.W.Hawking and G.F.Ellis, The Large Scale Structure of Spacetime
(Cambridge Univ. Press, Cambridge 1973).\par
[5] R. Penrose, Techniques of Differential Topology in Relativity,
(Society for Industrial and Applied Mathematics, Philadelphia 1972).\par
[6] E. Malec, Acta Phys. Pol. B22, 829(1991).\par
[7] J.Jezierski, J.Kijowski, Phys. Rev. D36, 1041(1987)\par
[8] P.Bizo\'n, E.Malec, N.O'Murchadha, Class. Quantum Grav.7,1953(1990) \par
[9] P.Koc, E.Malec, Acta Phys. Pol. B23, 123(1992); E.Flanagan, Phys. Rev. D46,
1429(1992).\par
[10] E. Malec, Acta Phys. Pol. B22, 347(1991).\par
[11] E. Malec, Phys. Rev. Lett. 67, 949(1991).\par
[12] P.Bizo\'n, E.Malec, N.O'Murchadha, Phys. Rev. Lett. 61, 1147(1988).\par
[13] E.Malec, Mod. Phys. Lett. A7,1679 (1992).\par
[14] D.M. Eardley, L.Smarr, Phys. Rev. D19, 2239(1979).\par
\end